\documentclass[aps,pre,showpacs,twocolumn,superscriptaddress]{revtex4}
\usepackage{graphicx,color,epsfig}
\begin{document}

\pdfoutput=1
\title{Threshold learning dynamics in social networks}

\author{J. C. Gonz\'alez-Avella}
\affiliation{Instituto de F\'isica Interdisciplinar y
Sistemas Complejos IFISC  (CSIC-UIB), E-07122 Palma de Mallorca, Spain}

\author{V. M. Egu\'iluz}
\affiliation{Instituto de F\'isica Interdisciplinar y
Sistemas Complejos IFISC (CSIC-UIB), E-07122 Palma de Mallorca, Spain}

\author{M. Marsili}
\affiliation{The Abdus Salam International Centre for Theoretical Physics, Trieste, Italy}

\author{F. Vega-Redondo}
\affiliation{European University Institute, Florence, Italy}
\affiliation{Instituto Valenciano de Investigaciones Econ\'omicas}

\author{M. San Miguel}
\affiliation{Instituto de F\'isica Interdisciplinar y
Sistemas Complejos IFISC (CSIC-UIB), E-07122 Palma de Mallorca, Spain}
\date{\today}

\begin{abstract}
Social learning is defined as the ability of a population to aggregate information, a process which must crucially depend on the mechanisms of social interaction. Consumers choosing which product to buy, or voters deciding which option to take respect to an important issues, typically confront external signals to the information gathered from their contacts. Received economic models typically predict that correct social learning occurs in large populations unless some individuals display unbounded influence. We challenge this conclusion by showing that an intuitive threshold process of individual adjustment does not always lead to such social learning. We find, specifically, that three generic regimes exist. And only in one of them, where the threshold is within a suitable intermediate range, the population learns the correct information. In the other two, where the threshold is either too high or too low, the system either freezes or enters into persistent flux, respectively. These regimes are generally observed in different social networks (both complex or regular), but limited interaction is found to promote correct learning by enlarging the parameter region where it occurs.
\end{abstract}

\maketitle

\section{Introduction}
Social learning has been a topic of central concern in economics during the
last decades \cite{Fudenberg98}, as it is central to a wide range of
socio-economic phenomena. Consumers who want to choose among a given set of
available products may seek the opinion of people they trust, in addition to
the information they gather from prices and/or advertisement. And voters who
have to decide what candidate to support in an election, or citizens who
have to take a stand on some issue of social relevance may rely on their
contacts to form their opinion. Ultimately, whether our societies take the
right course of action on any given issue (e.g. on climate change) will
hinge upon our ability to aggregate individual information that is largely
disperse. Thus, in particular, it must depend on the information diffusion mechanism by
which agents learn from each other, and therefore on the underlying social
network in which they are embedded. The significance of the conceptual
challenges raised by these issues is made even more compelling by the
booming advance in Information and Communication Technologies, with its
impact on the patterns of influence and communication, and on the way and
speed in which we communicate.

These key issues have attracted the interest of researchers in several
fields. For example, the celebrated ``voter model" \cite{Holloy75,Ligget85} is a prototype of those simple mechanistic models that are very
parsimonious in the description of individual behavior but allow for a full
characterization of the collective behavior induced. The voter model
embodies a situation where each agent switches to the opinion/state held by
one of the randomly selected neighbors at some given rate, and raises the
question of whether the population is able to reach consensus, i.e. a
situation where all agents display the same state. The literature on
consensus formation, as reviewed e.g. in Refs.~\cite{Castellano09,SanMiguel05}, has focused, in particular, on the role played
by the structure of the underlying network in shaping the asymptotic
behavior. One of the main insights obtained is that the higher is the
effective dimensionality of the network the harder it is for conformity to
obtain \cite{Suchecki05,Vazquez08}. Consensus formation in social systems is
closely related to the phenomenon of social learning. Indeed, the latter can
be regarded as a particular case of the former, when consensus is reached on
some ``true" (or objective) state of the
world, for example, given by an external signal \cite{Centola07a,GonzalezAvella10} impinging on the social
dynamics.

At the opposite end of the spectrum, economists have stressed the \emph{micro-motives} that underlie individual behavior and the assumption of
rationality. They have also emphasized the importance of going beyond models
of global interaction and/or bilateral random matching, accounting for some
local structure (modeled as a social network) in the pattern of influence
or communication among agents. This literature (see Ref.~\cite{Goyal05} for
an early survey) has considered a number of quite different scenarios,
ranging from those where agents just gather and refine information \cite{DeGroot74,Bala98,DeMarzo03} to contexts
where, in addition, there is genuine strategic interaction among agents \cite{Gale03}. Despite the wide range of specific models considered,
the literature largely conveys a striking conclusion: full social conformity
is attained (although not necessarily correct learning), irrespectively of
the network architecture. On the other hand, to attain correct learning, one
must require not only that the population be large but, in the limit, that
no individual retain too much influence \cite{Gale03,Golub09}.

The model studied in this paper displays some similarities to, as well as
crucial differences with, those outlined above. To fix ideas, the model could be regarded as reflecting a situation
where, despite the fact that new information keeps arriving throughout, the
consequences of any decision can only be observed in the future. Even more
concretely, this could apply, for example, to the performance of a political
candidate, the health consequences of consuming a particular good, or the
severity of the problem of climate change, on all of which a flow of fresh
information may be generated that is largely independent of agents' evolving
position on the issue. So, as in Ref.~\cite{GonzalezAvella10}, the agents receive an external signal; however, the
signal is noisy and it is confronted with the behavior displayed by
neighbors. As in Refs.~\cite{Bala98,Gale03}, while the agents make and
adjust their choices, they keep receiving noisy signals on what is the best
action. In contrast, however, these signals are not associated to
experimentation. In this respect, we share with Ref.~\cite{DeMarzo03,Golub09} the
assumption that agents' arrival of information is not tailored to current
choices.

The problem, of course, would become trivially uninteresting if agents either have unbounded memory or store information that is a
sufficient statistic for the whole past (e.g. updated beliefs in a Bayesian
setup). For, in this case, agents could eventually learn the best action by
relying on their own information alone. This leads us to making the stylized
assumption that the particular action currently adopted by each individual
is the only ``trace" she (and others) keep
of her past experience. Thus her ensuing behavior can only be affected by
the signal she receives and the range of behavior she observes (i.e., her
own as well as her neighbors'). Under these conditions, it is natural to
posit that if an agent receives a signal that suggests changing her current
action, she will look for evidence supporting this change in the behavior
she observes on the part of her neighbors. And then, only if a high enough
fraction of these are adopting the alternative action, she will undertake
the change. This, indeed, is the specific formulation of individual learning
studied in the present paper, which is in the spirit of the many threshold
models studied in the literature, such as \cite{Granovetter78,Morris00,Watts02,Centola07b}.

In the setup outlined, it is intuitive that the ``acceptance
threshold" that agents require to abandon the status quo
should play a key role in the overall dynamics. And, indeed, we find that
its effect is very sharp. First, note the obvious fact that if the threshold
is either very high or very low, social learning (or even behavioral
convergence) cannot possibly occur. For, in the first case (a very high
threshold), the initial social configuration must remain frozen, while in
the second case (a very low threshold), the social process would enter into
a state of persistent flux where agents keep changing their actions. In both
of these polar situations, therefore, the fraction of agents choosing the
good action would center around the probability $p$ with which the signal
favors that action.

Outside of these two polar situations, we find that there is always an
intermediate region where social learning does occur. And, within this
region, learning emerges abruptly. Specifically, there are upper and lower
bounds (dependent on $p$) such that, if the threshold lies within these
bounds, \emph{all} agents learn to play the good action while \emph{no}
learning \emph{at all }occurs if the threshold is outside that range. Thus
the three aforementioned regions are separated by sharp boundaries. A
similar abruptness in learning arises as one considers changes in $p$. In
this case, there is a lower bound on $p$ (which depends on the threshold)
such that, again, we have a binary situation (i.e., no learning or a
complete one) if the informativeness of the signal is respectively below or
above that bound. In a sense, these stark conclusions highlight the
importance of the social dimension in the learning process. They show that,
when matters/parameters are ``right," the
process of social learning builds upon itself to produce the sharp changes
just outlined.

As it turns out, this same qualitative behavior is encountered in a wide
variety of different network contexts. To understand the essential features
at work, we start our analysis by studying the simple case of a complete
graph, where every agent is linked to any other agent. This context allows
one to get a clear theoretical grasp of the phenomenon. In particular, it
allows us to characterize analytically the three different regimes of social
learning indicated: correct learning, frozen behavior, or persistent flux.
We then show that this characterization also provides a good qualitative
description of the situation when the interaction among agents is mediated
via a sparse complex network. We consider, in particular, three paradigmatic
classes of networks: regular two-dimensional lattices, Poisson random
networks, and Barab\'asi-Albert scale free networks. For all these cases, we
conduct numerical simulations and find a pattern analogous to the one
observed for the complete graph. The interesting additional observation is
that local interaction \emph{enlarges} (in contrast to global interaction)
the region where social learning occurs. In fact, this positive effect is
mitigated as the average degree of the network grows, suggesting a positive
role for relatively limited/local connectivity in furthering social learning.

\section{The model\label{Model}}

There is large population of agents, ${\cal N}=\{1,2,...,N\}$, placed on a given
undirected network $\Gamma =({\cal N},L)$, where we write $ij\in L$ if there is
link between nodes $i$ and $j$ in $\Gamma $. Let time step $s=0,1,2,...$ be
indexed discretely. Each agent $i\in {\cal N}$ displays, at any time step $s$, one of
two alternative actions $a_{i}(s)=\pm 1$, which are not equivalent. One of
them, say action $1$, induces a higher (expected) payoff, but the agents do
not know this.

At each time step $s$, one randomly chosen agent $i\in {\cal N}$ receives a signal on the
relative payoff of the two actions. This signal, which is independent across
time and agents, is only partially informative. Specifically, it provides
the correct information (i.e., ``action 1 is best") with
probability $p>1/2$, while it delivers the opposite information with the
complementary probability $1-p$.

If agent $i$'s previous action $a_{i}(s-1)$ does not coincide with the
action $\alpha_{i}(s)$ suggested as best, she considers whether changing to
the latter. We assume that she chooses $\alpha_{i}(s)$ (thus making $a_{i}(s)=\alpha_{i}(s)$) if, and only if, the fraction of neighbors in $\mathcal{N}_{i}\equiv \left\{ j\in {\cal N}:ij\in L\right\} $ who chose $\alpha_{i}(s)$ at $s-1$ exceeds a certain threshold. Let this (common) threshold
be denoted by $\tau \in \lbrack 0,1]$.

At the start of the dynamic process, each agent receives one signal $\alpha_{i}(0)$ and adopts the corresponding action. In other words, the initial
condition for the process is one where each agent, independently, holds
action $+1$ with probability $p$ or action $-1$ with probability $1-p$.

The central question posed in the paper can now be precisely formulated:
\begin{quote}
What is the relationship between $p$ (the quality of the signal) and $\tau $
(the threshold for action change) that underlies the spread and
consolidation of action $1$?
\end{quote}
This is the question addressed in what follows, in a range of different
setups and relying on a variety of methodologies.

\section{Global interaction for infinite populations \label{MFA}}

\begin{figure}[t]
\centerline{\includegraphics[width=.5\textwidth]{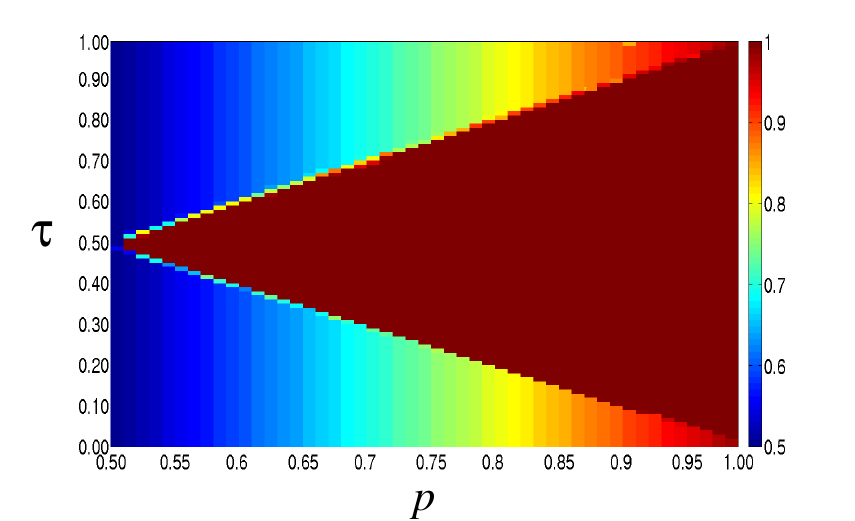}}
\caption{Phase diagram of the threshold model on a fully connected network.
The colors represent the fraction of agents choosing action $1$ (from red, $x=1$ to blue, $x=0.5$). System size given by $N=10^{4}$ agents; averaged over $100$ realizations.}
\label{Fig1}
\end{figure}

Let us consider the case where interaction is global: for each pair of
agents $i,j$ we have that $i\in \mathcal{N}_{j}$ and $j\in \mathcal{N}_{i}$.
Let $x(t)\in \lbrack 0,1]$ stand for the fraction of agents choosing action $1$ at time $t=s/N$. In the limit of infinite population size ($N\rightarrow
\infty $), the dynamics is given by:
\begin{equation}
\dot{x}=-(1-p)x~\theta (1-x-\tau )+p(1-x)~\theta (x-\tau )  \label{mfeq}
\end{equation}%
where $\theta (z)=1$ if $z\geq 0$ while $\theta (z)=0$ if $z<0$. This
equation is derived by considering the change $\mathrm{d}t$ in the fraction $x$ occurring in a time interval of $n\mathrm{d}t$ time steps. For $N\rightarrow \infty $, for any finite $\mathrm{d}t$, this increment
converges, by the law of large numbers, to a constant given by the right
hand side of Eq.~(\ref{mfeq}) times $\mathrm{d}t$. The first term accounts
for the number of agents initially with the right signal ($x$) who receive
the wrong signal (with probability $1-p$) and adopt it, as the fraction of
agents also adopting it is larger than the threshold ($1-x>\tau $). The
second accounts for the opposite subprocess, whereby agents who receive the
correct signal (with probability $p$) switch to the correct action when the
population supports it ($x>\tau $).

We assume that, at time $t=0$, each agent receives a signal $\alpha_i(0)$
and adopts the corresponding action $a_i(0)=\alpha_i(0)$. Hence the initial
condition for the dynamics above is $x(0)=p$.

It is useful to divide the analysis into two cases:

\begin{description}
\item[Case I: $\tau >1/2$]
\end{description}
In this case, it is straightforward to check that
\[
\begin{array}{ccc}
x<1-\tau & \Longrightarrow & \dot{x}=-(1-p)x<0 \\
1-\tau <x<\tau & \Longrightarrow & \dot{x}=0 \\
x>\tau & \Longrightarrow & \dot{x}=p(1-x)>0%
\end{array}%
\]%
\noindent So, it follows that correct social learning occurs iff $p>\tau $.

\begin{description}
\item[Case II: $\tau <1/2$]
\end{description}
In this case, we find:
\[
\begin{array}{ccc}
x<\tau & \Longrightarrow & \dot{x}=-(1-p)x<0 \\
\tau <x<1-\tau & \Longrightarrow & \dot{x}=p-x \\
x>1-\tau & \Longrightarrow & \dot{x}=p(1-x)>0%
\end{array}%
\]%
\noindent And, therefore, correct social learning occurs iff $p>1-\tau $.

\begin{figure}[t]
\centerline{\includegraphics[width=.5\textwidth]{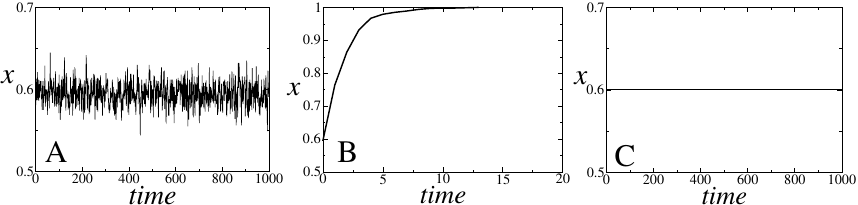}}
\caption{Typical realizations of the time evolution of the fraction of agents choosing action 1, $x$, in a fully connected network of system size $N=10^4$with $p=0.60$, and ({\it A}) $\tau =0.20$; ({\it B}) $\tau =0.50$; $({\it C})$ $\tau =0.80$.}
\label{Fig2}
\end{figure}

Combining both cases, we can simply state that, in the global interaction
case, correct social learning, $x_{\infty }=x(t\rightarrow \infty )=1$
occurs if, and only if,
\begin{equation}
\tau \in (1-p,p)~,  \label{range-1}
\end{equation}
that is, the threshold $\tau $ is within an intermediate region whose size
grows with the probability $p$, which captures the informativeness of the
signal. However, there are other two phases: if $\tau \in (0,1-p)$, the
system reaches the stationary solution $x_{\infty }=p$; while if $\tau \in
(p,1)$, we have $\dot{x}=0$ for all times, which means that the system stays
in the initial condition $x_{\infty }=x(t=0)=p$.

\begin{figure}[t]
\centerline{\includegraphics[width=.5\textwidth]{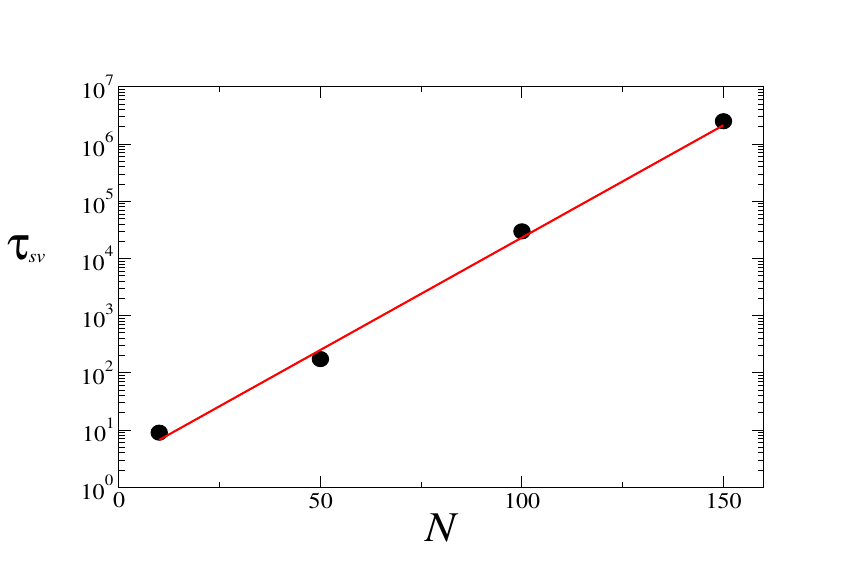}}
\caption{The average survival time $\tau_\textrm{sv}$ in fully connected networks for
different system sizes $N$ for $p=0.60$ and $\tau =0.20$. The
continuous line corresponds to an exponential fit of the form $\tau_\textrm{sv}\sim \exp^{(c N)}$, being $c$ a constant.}
\label{Fig3}
\end{figure}

\section{Numerical simulations}

Now we explore whether the insights obtained from the infinite size limit of
the global interaction case carry over to setups with a finite but large
population, where agents are genuinely connected through a social network.

First, we consider the benchmark case of global interaction (i.e., a
completely connected network). Then, we turn to the case of local
interaction and focus on three paradigmatic network setups: lattice
networks, Erd\"{o}s-R\'enyi (Poisson) networks, and Barab\'{a}si-Albert
(scale-free) networks.

\subsection{Global interaction\label{FullyConnected}}

The results obtained on the completely connected network (i.e., the network
where every pair of nodes is linked) are in line with the theory presented
in the previous section. The essential conclusions can be summarized through
the phase diagram in the $(p,\tau )$-space of parameters depicted in Figure~1. There we represent the fraction of agents choosing action $1$ in
the steady state for each parameter configuration, with the red color
standing for a homogeneous situation with $x=1$ (i.e., all agents choosing
action $1$) while the blue color codes for a situation where $x=0.5$ and
therefore the two actions are equally present in the population.
Intermediate situations appear as a continuous color grading between these
two polar configurations.

We find that, depending on the quality of the external signal $p$ and the
threshold $\tau $, the system reaches configurations where either complete
learning occurs ($x=1$) or not ($x=p$). Indeed, the observed asymptotic
behavior is exactly as predicted by the analysis of the previous section and
it displays the following three phases:

\begin{itemize}
\item Phase I: $\tau <1-p$. The system reaches a stationary aggregate
configuration where the nodes are continuously changing their state but the
average fraction of those choosing action $1$ gravitates around the
frequency $x=p,$ with some fluctuations (see Figure~2{\it A}). The
magnitude of these fluctuations decreases with system size $N$.

\item Phase II: $1-p<\tau <p$. The system reaches the absorbing state $x=1$
where everyone adopts action $1$. This is a situation where the whole
population eventually learns that the correct choice is action $1$ (see
Figure~2{\it B}).

\item Phase III: $\tau >p$. The system freezes in the initial state, so the
fraction $x=p$ of agents choosing the correct action coincides with the
fraction of those that received the corresponding signal at the start of the
process (see Figure~2{\it C}).
\end{itemize}

It is worth noting that, while in Phase I the theory predicts $x=p$, any
finite-size system must eventually reach an absorbing homogenous state due
to fluctuations. Thus, to understand the nature of the dynamics, we
determine the average time $\tau_\textrm{sv}$ that the system requires to reach
such an absorbing state. As shown in Figure~3, $\tau_\textrm{sv}$
grows exponentially with $N$. This means that $\tau_\textrm{sv}$ grows very fast
with system size, and thus the coexistence predicted by the theory in Phase
I can be regarded as a good account of the situation even when $N$ is just
moderately large.

\subsection{Lattice networks}

Now assume that all nodes are placed on a \emph{regular boundariless lattice
}of dimension $2$, endowed with the distance function $d(\cdot )$ given by $%
d(i,j)=\max \left\{ \left\vert x_{i}-x_{j}\right\vert ,\left\vert
y_{i}-y_{j}\right\vert \right\} $. The social network is then constructed by
establishing a link between every pair of agents lying at a lattice distance
no higher than some pre-specified level $\bar{d}$. This defines the
neighborhood $\mathcal{N}_{i}$ of any agent $i\in N$, as given by $\mathcal{N%
}_{i}=\{j\in N:d(j,i)\leq \bar{d}\}$. In this network, the degree (i.e. the
number of neighbors) of any node $k_{i}=k$ is related to $\bar{d}$; for
instance, if $\bar{d}=1$ we have $k=8$.

\begin{figure}[t]
\centerline{\includegraphics[width=.4\textwidth]{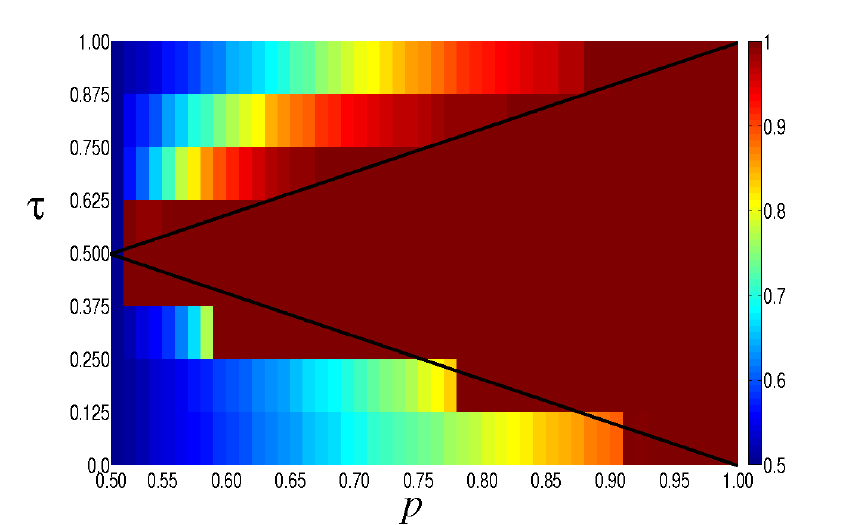}}
\caption{Phase diagram of the threshold model on a two-dimensional lattice
with $k=8$ ($\bar{d}= 1$). The colors represent the
fraction of agents choosing action 1action (from red, $x=1$, to blue, $x=0.5$). System size $N=10^{4}$; average over $100$ realizations.}
\label{Fig4}
\end{figure}

The behavior of the system is qualitatively similar to the case of a fully
connected network. Again we find three phases. In two of them, both actions
coexist with respective frequencies $p$ and $1-p$ (one phase is frozen and
the other continuously fluctuating), while in another one the whole
population converges to action $1$. A global picture of the situation for
the entire range of parameter values is shown in Figure~4, with the
black diagonal lines in it defining the boundaries of the full-convergence
region under global interaction. In comparison with the situation depicted
in Figure~1, we observe that the region in the $(p,\tau )$-space
where behavioral convergence obtains in the lattice network is broader than
in the completely connected network. This indicates that restricted (or
local) interaction facilitate social learning, in the sense of enlarging the
range of conditions under which the behavior of the population converges to
action $1$.

As a useful complement to the previous discussion, Figure~5
illustrates the evolution of the spatial configuration for a typical
simulation of the model in a lattice network, with different values of $\tau$ and $p=0.6$. Panels $a$, $b$ and $c$ show the configurations of the system
for a low value of $\tau =1/8$ at three different time steps: $t=0$, $1000$
and $2000$ respectively. The evolution of the system displays a
configuration analogous to the initial condition, both actions coexisting
and evenly spreading throughout the network. This is a situation that leads
to dynamics of the sort encountered in \textit{Phase I} above. In contrast,
Panels $g$, $h$ and $i$ correspond to a context with a high $\tau =7/8$,
which induces the same performance as in \textit{Phase III}. It is worth
emphasizing that although Panels $a$, $b$ and $c$ display a similar spatial
pattern, they reflect very different dynamics, i.e., continuous turnover in
the first case, while static (frozen initial conditions) in the second case.
Finally, Panels $d$, $e$ and $f$ illustrate the dynamics for an intermediate
value of $\tau =1/2$, which leads to a behavior of the kind displayed in
\textit{Phase II}. Specifically, these panels show that, as the system moves
across the three time steps: $t=0$, $16$ and $21$, the system evolves, very
quickly, toward a state where all agents converge to action $1$.

\begin{figure}[t]
\centerline{\includegraphics[width=.4\textwidth]{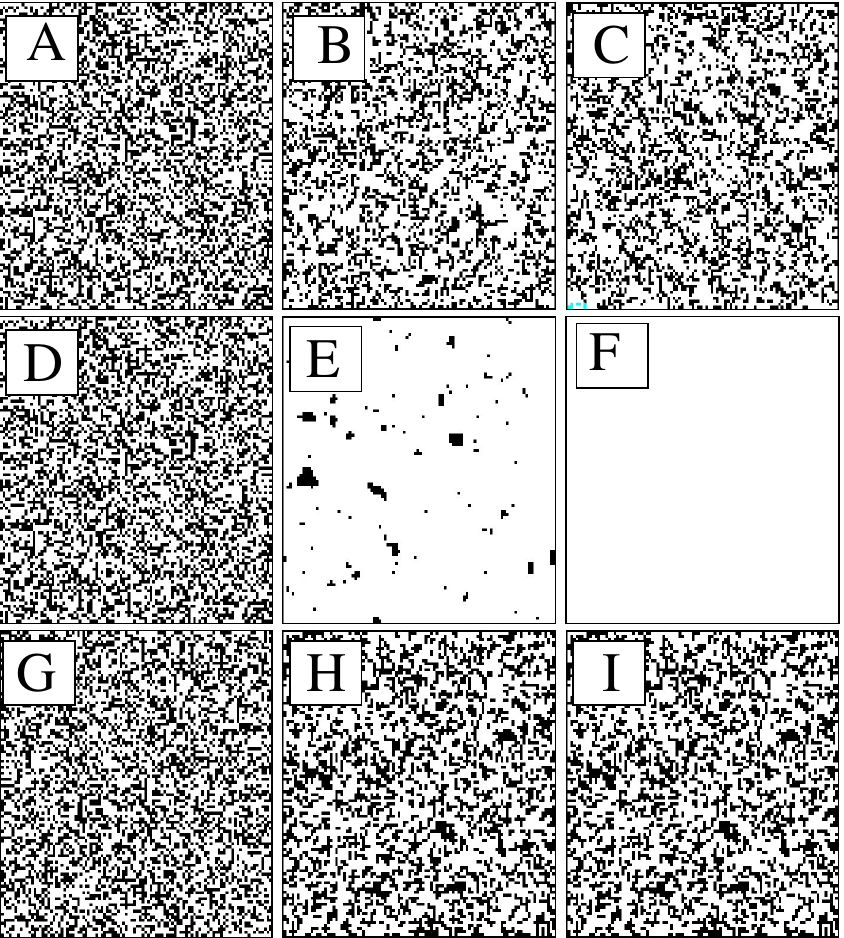}}
\caption{Time evolution of the threshold model on a two-dimensional lattice
with $k=8$ for different values of $\tau $ and $p=0.60$. Panels ({\it A-C}): $\tau =\frac{1}{8}$ and time steps ({\it A}) $t=0$, ({\it B}) $1000$ and ({\it C}) $2000$. Panels ({\it D-F}): $\tau =\frac{1}{2}$ and time steps ({\it D}) $t=0$, ({\it E}) $16$ and ({\it F}) $t_{3}=21$. Panels ({\it G-I}): $\tau =\frac{7}{8}$ and time steps ({\it G}) $t=0$, ({\it H}) $1000$ and ({\it I}) $2000$. Black color represents an agent using action $-1$, while white color represents action $1$. The system size is $N=10^{4}$.}
\label{Fig5}
\end{figure}

\subsection{Erd\"{o}s-R\'enyi and scale-free networks}

A lattice network is the simplest possible context where local interaction
can be studied. It is, in particular, a regular network where every agent
faces exactly symmetric conditions. It is therefore interesting to explore
whether any deviation from this rigid framework can affect our former
conclusions. This we do here by focusing on two of the canonical models
studied in the network literature: the early model of Erd\"{o}s and R\'{e}nyi (ER) \cite{Erdos59} and the more recent scale-free model introduced by
Barab\'{a}si and Albert (BA) \cite{Barabasi99}. Both of them abandon the
regularity displayed by the lattice network and contemplate a non-degenerate
distribution of node degrees.

The ER random graph is characterized by a parameter $\mu $, which is the
connection probability of agents. It is assumed, specifically, that each
possible link is established in a stochastically independent manner with
probability $\mu $. Consequently, for any given node, its degree
distribution $P\equiv \{P(k)\}$ determining the probability that its degree
is $k$ is Binomial, i.e., $P(k)={N-1 \choose k} \mu ^{k}(1-\mu )^{N-1-k}$,
with an expected degree given by $\left\langle k\right\rangle =\mu (N-1)$.
In the simulations reported below, we have focused on networks with $\langle k \rangle =8$ and $N=10^{4}$.

On the other hand, to build a BA network, we follow the procedure described
in Ref.~\cite{Barabasi99}. At each time step, a new node is added to the
network and establishes $m$ links to existing nodes. The newcomer selects
its neighbors randomly, with the probability of attaching to each of the
existing nodes being proportional to their degree $k$. It is well known that
this procedure generates networks whose degree distribution follows a power
law of the form $P(k)\simeq 2m^{2}k^{-\gamma }$, with $\gamma \approx 3$.
For our simulations, we have constructed BA networks using this procedure
and a value of $m=4$, leading to an average degree $\langle k\rangle =2m=8$.

\begin{figure}[t]
\centerline{\includegraphics[width=.4\textwidth]{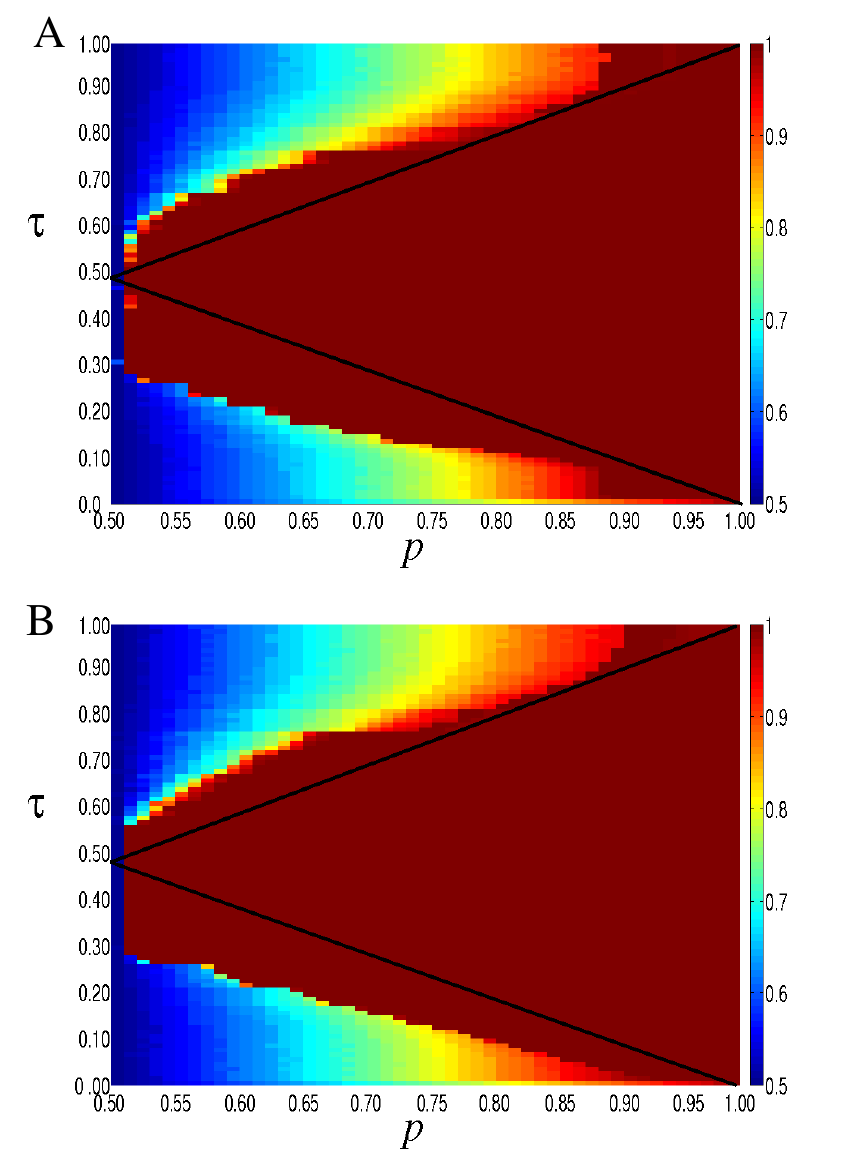}}
\caption{Phase diagram of the threshold model in a ({\it A}) ER network and in a
({\it B}) scale-free network with average degree $\langle k \rangle = 8$. The
colors represent the fraction of agents choosing action 1 (from red, $x=1$,
to blue $x=0.5$). System size $N=10^4$, average over $100$ realizations.}
\label{Fig6}
\end{figure}

The networks are constructed, therefore, so that they have the same average
degree in both the ER and BA contexts. It is important to emphasize,
however, that the degree distributions obtained in each case are markedly
different. While in the former case, the degree distribution induces an
exponentially decaying probability for high-degree nodes, in the latter case
it leads to ``fat tails", i.e. associates significant
probability to high-degree nodes.

The results are illustrated in Figure~6. For the two alternative
network topologies, the system displays qualitatively the same behavior
found in the lattice network. That is, there are three distinct phases
yielding distinct kinds of dynamic performance: convergence to action $1$,
frozen behavior, and persistent turnover. However, it is interesting to note
that, compared with the case of global interaction, the convergence region
(which we labeled as Phase II before) is significantly larger. This suggests
that local (i.e. limited) connectivity facilitates social learning.

Why does limited connectivity extend the learning region? Intuitively, the
reason is that it enhances the positive role in learning played by random
fluctuations. Such fluctuations are neglected, by construction, in the
mean-field approximation and are also minimized when the whole population
interacts globally. But, when interaction is local, those fluctuations will
tend to de-stabilize the situation in both the constant flux and in the
frozen phases -- at first, locally, but then also globally.

To gain a more refined understanding of this issue, let us try to assess the
effect of local interaction on the likelihood that, at some random initial
conditions, any given node faces a set of neighbors who favors a change of
actions. This, of course, is just equal to the probability that the fraction
of neighbors who display opposite behavior is higher than $\tau $, the
required threshold for change. Thus, more generally, we want to focus on the
conditional distribution densities $\phi_{+}(\nu )$ and $\phi_{-}(\nu )$ that
specify, for an agent displaying actions $1$ and $-1$ respectively, the
probability density of finding a fraction $\nu$ of neighbors who adopt actions $-1$
and $1$, respectively. Of course, these distributions must depend on the
degree distribution of the network and, in particular, on its average
degree. Specifically, when the average degree of the network is large
relative to population size (thus we approximate a situation of global
interaction) those distributions must be highly concentrated around $p$ and $%
1-p$ respectively. Instead, under lower connectivity (and genuine local
interaction), the distributions $\phi_{+}(\nu )$ and $\phi_{-}(\nu )$ will
tend to be quite disperse.

Next, let us understand what are the implications of each situation. In the
first case, when the connectivity is high, the situation is essentially
captured by a mean-field approximation, and thus the induced dynamics must
be well described by the global interaction case (in particular, as it
concerns the size of the convergence region). In contrast, when the
connectivity is low and the distributions $\phi_{+}(\nu )$ and $\phi_{-}(\nu )$
are disperse, a significant deviation from the mean-field theory is
introduced. In fact, the nature of this deviation is different depending on
the level of the threshold $\tau $. If it is low, and thus action turnover
high, it mitigates such turnover by increasing the probability that the
fraction of neighbors with opposite behavior lie below $\tau $. Instead, if $\tau $ is high and action change is difficult, it renders it easier by
increasing the probability that the fraction of neighbors with opposite
behavior lies above $\tau $. Thus, in both cases it works against the forces
that hamper social learning and thus improves the chances that it occurs.

More precisely, the above considerations are illustrated in Figure~7 for a lattice network. There we plot the distributions $\phi_{+}(\nu )$ for different levels of connectivity $k$ and parameter values $p=0.60$ and $\tau =0.30$ -- recall that these values correspond to Phase I
(with high turnover) in a \emph{fully connected} network. Consider first the
situation that arises for values of $k=8,~24,~56$ -- i.e. low connectivity
relative to the size of the system. Then we find that, among the nodes that
are adopting action $1$, $\phi_{+}$ attributes a significant probability
mass to those agents whose fraction of neighbors $\nu $ choosing action $-1$
is below the threshold required to change (as marked by the vertical dashed
line). Such nodes, therefore, will not change their action. And, as
explained, this has the beneficial effect of limiting the extent of action
turnover as compared with the global interaction setup. On the other hand,
the inset of Figure~7 shows that, among the nodes that are
adopting action $-1$, the distribution $\phi_{-}$ associates a large
probability mass to those agents whose fraction of neighbors $\nu$ choosing
the opposite action is above $\tau $. This ensures that there is a large
enough flow from action $-1$ to action $1$. In conjunction, these two
considerations lead to a situation that allows, first, for some limited
nucleation around action $1$ to take place, followed by the ensuing spread
of this action across the whole system (Figure~7({\it B-D})).

Let us now reconsider the former line of reasoning when $k$ is large -- in
particular, take the case $k=828$ depicted in Figure~7. Then, the
corresponding distribution $\phi_{+}$ is highly concentrated around $\nu =p$, essentially all its probability mass associated to values that lie above $\tau =0.30$. This means that the induced dynamics must be similar to that
resulting from the complete-network setups, and thus too-fast turnover in
action choice prevents the attainment of social learning. Clearly, social
learning would also fail to occur for such high value of $k$ if the
threshold $\tau$ were large. In this case, however, the problem would be
that the highly concentrated distributions $\phi_{+}$ and $\phi_{-}$ would
have most of their probability mass lying below the threshold. This, in
turn, would lead to the freezing of the initial conditions, which again is
the behavior encountered for a complete network.

\begin{figure}[t]
\centerline{\includegraphics[width=.5\textwidth]{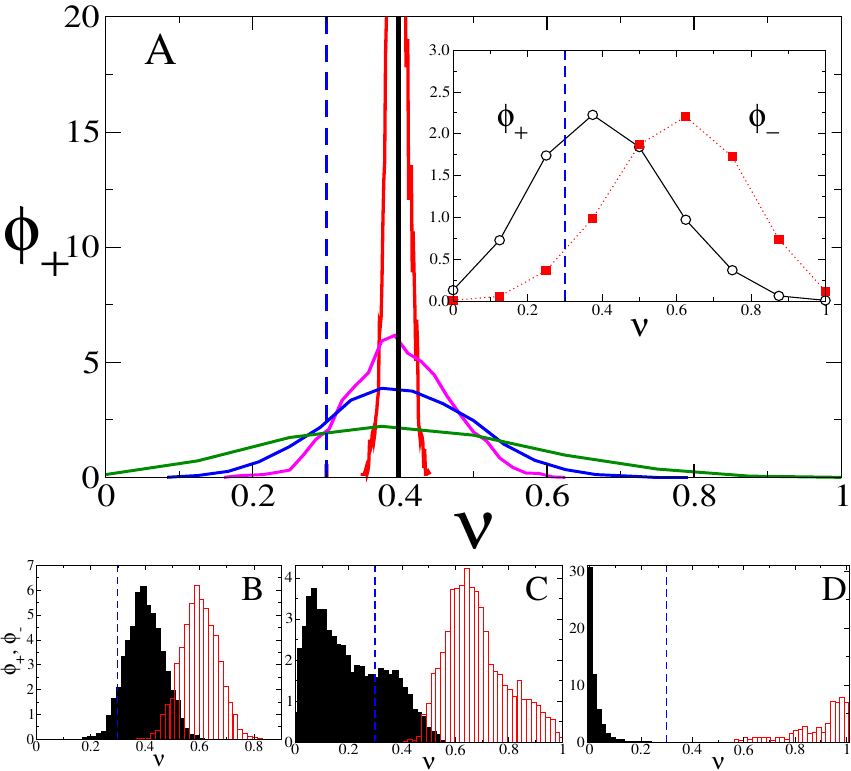}}
\caption{{{\it A}} The initial probability density $\phi_{+}$
that a node using action $1$ has a fraction of neighbor nodes with action $-1$, computed on a two-dimensional lattice for $k=8$, $24$, $56$, $828$ and a completely connected network (from the broadest to the narrowest probability density distribution). [\emph{Inset}: $\phi_{+}$ (black, continuous)
and $\phi_{-}$ (red, dotted) for $k=8$.] Time
evolution of the probability densities $\phi_{+}$ (black) and $\phi_{-}$ (red) in a two-dimensional lattice with $k=56$ for ({\it B}) $t=0$, ({\it C}) 5 and ({\it D}) 10. For all panels, the dashed line indicates the threshold $\tau =0.3$; parameter values: system size is $N=10^{4}$, $p=0.60$, and $\tau =0.30$. }
\label{Fig7}
\end{figure}


\section{Summary}

The paper has studied a simple model of social learning with the following
features. Recurrently, agents receive an external (informative) signal on
the relative merits of two actions. And, in that event, they switch to the
action supported by the signal if, and only if, they find support for it
among their peers - specifically, iff the fraction of these choosing that
action lies above a certain threshold. Given the quality of the signal,
correct social learning occurs iff the threshold is within some intermediate
region, i.e. neither too high nor too low. For, if it is too high, the
situation freezes at the configuration shaped at the beginning of the
process; and if it is too low, the social dynamics enters into a process of
continuous action turnover. A key conclusion is that social learning is a
dichotomic phenomenon, i.e. it either occurs completely or not at all,
depending on whether the threshold lies within or outside the aforementioned
region.

These same qualitative conclusions are obtained, analytically, in the case of global interaction --
which corresponds to a mean-field version of the model -- as well as, numerically, in a wide range of
social networks: complete graphs, regular lattices, Poisson random networks,
and Barab\'{a}si-Albert scale-free networks. However, the size of the
parameter region where social learning occurs depends on the pattern of
social interaction. In general, an interesting finding is that learning is
enhanced (i.e. the size of the region enlarged) the less widespread is such
interaction. This happens because genuinely local interaction favors a
process of spatial nucleation and consolidation around the correct action,
which can then spread to the whole population.

In sum, a central point that transpires from our work is that, in contrast
to what most of the received socio-economic literature suggests, social
learning is hardly a forgone conclusion. This, of course, is in line with
the common wisdom that, paraphrasing a usual phrase, crowds are not always
wise. In our threshold framework, this insight is robust to the topology or
density of social interaction. Furthermore, our results highlight the importance of identifying the information diffusion mechanism, and the local sampling of the population provided by the social network. But future research should explore whether it
is also robust to a number of important extensions. Just to mention a few,
these should include (a) interagent heterogeneity -- e.g. in their
individual thresholds for change; (b) different behavioral rules -- e.g.
payoff-based imitation; or (c) the possibility that agents adjust their
links, so that learning co-evolves with the social network.


\begin{acknowledgments}
This work was partially supported by MEC (Spain) through project FISICOS (FIS2007-60327).
\end{acknowledgments}


\begin{thebibliography}{99}
\bibitem{Fudenberg98} D. Fudenberg and D. K. Levine.  \emph{The Theory
of Learning in Games}, Cambridge: MIT Press, (1998).

\bibitem{Holloy75} R. Holley and T. Liggett.  Ergodic Theorems for Weakly Interacting Infinite Systems and the Voter Model, \textit{Annals of Probability \textbf{3}}, 643, (1975).

\bibitem{Ligget85} T.M. Liggett.  \emph{Interacting Particle Systems},
New York: Springer, (1985).

\bibitem{Castellano09} C. Castellano, S. Fortunato, and V. Loreto. 
Statistical physics of social dynamics, \textit{Rev. Mod. Phys. \textbf{81}}, 591, (2009).

\bibitem{SanMiguel05} M. San Miguel, V.M. Egu\'iluz, R. Toral, and K. Klemm.
Binary and Multivariate Stochastic models of consensus formation,
\textit{Computing in Science and Engineering \textbf{7}}, 67-73, (2005).

\bibitem{Suchecki05} K. Suchecki, V. M. Egu\'{\i}luz, and M. San Miguel.
 Voter model dynamics in complex networks: role of dimensionality,
\textit{Physical Review E \textbf{72}}, 036132, 1-8, (2005).

\bibitem{Vazquez08} F Vazquez and V. M. Egu\'{\i}luz.  Analytical
solution of the voter model on uncorrelated networks, \textit{New Journal of
Physics \textbf{10}}, 063011, (2008).

\bibitem{Centola07a} D. Centola, J.C. Gonz\'alez-Avella, V.M. Egu\'iluz, and
M. San Miguel.  Homophily, Cultural Drift, and the Co-Evolution of
Cultural Groups, \textit{Journal of Conflict Resolution \textbf{51}},
905-929, (2007).

\bibitem{GonzalezAvella10} J.C. Gonz\'alez-Avella, M.G. Cosenza, V.M.
Egu\'iluz, and M. San Miguel.  Spontaneous ordering against an external
field in nonequilibrium systems, \textit{New Journal of Physics \textbf{12}}, 013010, (2010).

\bibitem{Goyal05} S. Goyal. Learning in networks, in G. Demange and M.
H. Wooders (eds.), \emph{Group Formation in Economics: Networks, Clubs, and
Coalitions}, Cambridge: Cambridge University Press, (2005).

\bibitem{DeGroot74} M.H. De Groot. Reaching a Consensus, \textit{Journal of the American Statistical Association \textbf{69}}, 118-121, (1974).

\bibitem{Bala98} V Bala and S. Goyal. Learning from neighbours.
\textit{Review of Economic Studies \textbf{65}}, 595-621,  (1998).

\bibitem{Gale03} D. Gale and S. Kariv Bayesian learning in social
networks, \textit{Games and Economic Behavior \textbf{45}}, 329-346, (2003).

\bibitem{DeMarzo03} P. DeMarzo, D. Vayanos, and J. Zwiebel  Persuasion
bias, social influence, and unidimensional opinions, \textit{Quarterly
Journal of Economics \textbf{118}}, 909-968, (2003).

\bibitem{Golub09} B. Golub and M. Jackson  Naive learning in social
networks: convergence, influence and the wisdom of crowds, forthcoming in
\textit{American Economic Journal: Microecnomics,} (2009).

\bibitem{Granovetter78} M. Granovetter  Threshold Models of Collective
Behavior. \textit{American Journal of Sociology \textbf{83}}, 1420-1443, (1978).

\bibitem{Morris00} S. Morris Contagion, \textit{Review of Economic
Studies \textbf{67}}, 57 - 78, (2000).

\bibitem{Watts02} D.J. Watts. A simple model of global cascades on
random networks, \textit{Proc. Natll. Acad. Sci. USA \textbf{99}}, 5766-5771, (2002).

\bibitem{Centola07b} D. Centola, V.M. Egu\'iluz, and M.W. Macy 
Cascade dynamics of complex propagation, \textit{Physica A \textbf{374}},
449-456, (2007).

\bibitem{Erdos59} P. Erd\"os and A. R\'enyi  \textit{Publ. Math.
Debrecen \textbf{6}}, 290, (1959).

\bibitem{Barabasi99} A.-L. Barab\'asi and Albert R. \textit{Science
\textbf{286}}, 509, (1999).
\end{thebibliography}
\end{document}